\documentclass[multphys,vecphys]{svmult}

\usepackage{makeidx}   
\usepackage{graphicx}  
\usepackage{multicol}  

\makeindex             

\begin{document}
\title*{
Secular Decrease and  Random Variations of  Cassiopeia~A
at 151.5 and 927 MHz
}
\toctitle{
Secular Decrease and  Random Variations of  Cassiopeia~A
at 151.5 and 927 MHz
}
\titlerunning{Secular Decrease  of  Cassiopeia~A
at 151.5 and 927 MHz} 
\author{E.\,N.\,Vinyajkin, V.\,A.\,Razin}
\authorrunning{E.\,N.\,Vinyajkin, V.\,A.\,Razin}
\institute{Radiophysical Research Institute (NIRFI), 25 B.\,Pecherskaya st.,
Nizhny Novgorod,  603950, Russia\\
\texttt{evin@nirfi.sci-nnov.ru \ \ razin@nirfi.sci-nnov.ru}}

\maketitle              

\begin{abstract}
Long-term measurements of the radio flux density of Cassiopeia~A
relative to Cygnus~A have been carried out at 927 and 151.5~MHz. It was
found the following mean secular decrease rates of the radio emission of
Cassiopeia~A:
$(0.72\pm 0.03)\%\,\hbox{year}^{-1}$ at 927 MHz (for the period
1977--2002) and $(0.88\pm 0.09)\%\,\hbox{year}^{-1}$ at 151.5~MHz (for
the period 1980--2002). These values of the secular decrease rate
obtained over the period of the last 25 years are substantially less
than those of Baars et al. (1977). This indicates to the slowing down of
Cassiopeia~A radio emission secular decrease. In addition to this large
scale time variation of Cassiopeia~A flux density the measurements have
also shown a small scale (a few years) time variations over the smooth
secular decrease.
\end{abstract}

\section{ Introduction}
\label{sec1}

The secular decrease rate $d=S^{-1}dS/dt$ of the radio emission of young
supernova remnant Cassiopeia A was determined in many early
investigations using relatively few measurements (sometimes only
2--3 measurements) of its flux density $S$ at different epochs.
However, if we want to measure not only some mean $d$ over a long
time interval but also to reveal some possible time variations
of $S$, we have to make
more measurements. In addition, measurements at a given
frequency $\nu$ should be carried out using the same or rather similar
radio telescopes and identical measurement procedures.

This report presents the results of long-term (1977--2002)
measurements of the flux density of Cassiopeia~A relative to that of
Cygnus~A at 927 and 151.5~MHz.

\section{
Measurements of the Cassiopeia~A radio flux density
relative to Cygnus~A at 927 MHz
}
\label{sec2}

In the very begining of August 2002 we carried out the measurements of the Cassiopeia~A
radio flux density relative to Cygnus~A at the Radio Astronomical
Observatory ``Staraya Pustyn$'$" (geographical latitude
$55^\circ 39'$, longitude $2^{\rm h}54.5^{\rm m}$) using 10-m radio
telescope at 927~MHz. These measurements are an extension of the
long-term ones initiated in 1977~[1--3]. Using one and the same radio
telescope makes it possible to obtain a uniform observational material.
The measurement method consisted as before of successive registration
of Cassiopeia~A and Cygnus~A radio emission relative to definite
reference areas. One record of Cygnus~A had the following sequence
of antenna pointings lasted totally 6~minutes: reference
area\,--\,source\,--\,reference area
(``off"\,--\,``on"\,--\,``off"). The sequence of the same duration
for Cassiopeia~A was the following: first reference
area\,--\,source~--- second reference area
(``off1"\,--\,``on"\,--\,``off2").
The reference areas for Cassiopeia~A have the following coordinates:
right ascension
$\alpha_{\rm off1}=\alpha_{\rm CasA}-0^{\rm h}40^{\rm m}$,
$\alpha_{\rm off2}=\alpha_{\rm CasA}+0^{\rm h}40^{\rm m}$, declination
$\delta_{\rm off1;\,2}=\delta_{\rm CasA}$, that one for Cygnus~A has,
respectively: $\alpha_{\rm off}=20^{\rm h}12^{\rm m}$,
$\delta_{\rm off}=45^\circ05'$ (coordinates for the epoch 1950.0).
The radio emission of sources was registered at such time intervals
when the elevation difference of Cassiopeia~A and Cygnus~A by its
absolute value did not exceed $7^\circ$ at an average elevation of
both the sources $72^\circ$. These conditions define the
time and duration (about 2 hours) of one session of measurements.
As a result of two sessions we obtained the
following ratio of Cassiopeia~A and Cygnus~A flux densities at 927~MHz:

$${(S_{\rm CasA}/ S_{\rm CygA})}_{927\,{\rm MHz}}=1.096\pm0.011\hbox{ \
for the epoch 2002.58.}$$

\begin{figure}[h]
\begin{minipage}{60mm}
\includegraphics[width=6.01cm]{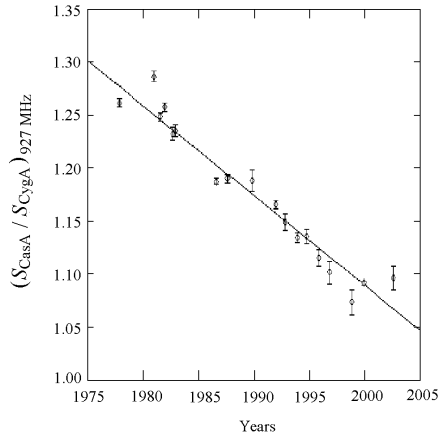}
\end{minipage}
\begin{minipage}{40mm}
\caption[]{Flux density of the Cassiopeia A radio emission
relative to that of Cygnus~A at 927~MHz versus time.} \label{F1}
\end{minipage}
\end{figure}

Figure 1 shows all values of
${(S_{\rm CasA}/ S_{\rm CygA})}_{927\,{\rm MHz}}$ for all years of
measurements using RT-10 at 927~MHz [1--3] obtained from measured
values by multiplying by 0.89 to take into account the difference of
brightness temperatures in the direction of Cygnus~A and its reference
area.

\section{
Interferometric measurements of the Cassiopeia~A radio
flux density relative to Cygnus~A at 151.5 MHz
}
\label{sec3}

In August -- September 2002 we carried out the measurements  of
Cassiopeia~A radio flux density relative to Cygnus~A at the Radio
Astronomical Observatory ``Staraya Pustyn$'$" using the
interferometer consisting of two 14-m radio telescopes at 151.5~MHz.
One session of measurements consisted of a
one hour record of Cygnus~A fringes near the upper culmination,
then that of Cassiopeia~A near also the upper culmination
and calibrations by the noise generator.
For this interferometer with a base of $31\lambda$ both sources
are practically point ones. By each measurement session we defined the
amplitude ratio of fringes of Cassiopeia~A and Cygnus~A equal to the
ratio of their flux densities. There were five measurement sessions. As a
result we got:
$$
{(S_{\rm CasA}/ S_{\rm CygA})}_{151.5\,{\rm MHz}}=0.91\pm0.01 \hbox{ \
for the epoch 2002.67.}$$

Figure 2 shows the values of ${(S_{\rm CasA}/S_{\rm
CygA})}_{151.5\,{\rm MHz}}$ for all years of measurements using
the interferometer RT-14$+$RT-14(2) at ``Staraya Pustyn$'$" at
151.5~MHz [2, 3].

\section{
Analysis of the results of long-term measurements\newline of
Cassiopeia~A flux densities at 927 MHz
}
\label{sec4}

Figure 1 shows the measurement results of the Cassiopeia~A radio flux
density relative to Cygnus~A at 927~MHz
${(S_{\rm CasA}/S_{\rm CygA})}_{927\,{\rm MHz}}\equiv r_{927}(t)$
made during 25 years (1977--2002) using one and the same 10-m radio
telescope at the NIRFI Radio Astronomical Observatory ``Staraya
Pustyn$'$". The straight line of Fig.\,1
$$
r_{927}(t) = m_{927}(t-\langle t\rangle_{927}) + c_{927},$$
where

${\langle t\rangle}_{927} = 1990.2$  is the mean epoch of
measurements at 927~MHz,

$
m_{927}  = dr_{927}(t) /dt =-(8.467 \pm 0.390)\cdot 10^{-3}
\hbox{\,year}^{-1},$

$
c_{927} = r_{927}(\langle t\rangle_{927}) = 1.172 \pm 0.003,$\\
shows a weighted least-squares fit.

The average value of the secular decrease rate of the Cassiopeia~A radio
emission over the time interval 1977--2002 is equal to
\begin{equation}\label{f1}
d_{927\,{\rm MHz}}(1977\!-\!2002)=
100\cdot m_{927}/c_{927} =- (0.72 \pm 0.03)\% \hbox{\,year}^{-1}.
\end{equation}

\section{
Analysis of the results of long-term measurements of
Cassiopeia A flux densities at 151.5 MHz
}
\label{sec5}

Figure 2 shows the measurement results of the Cassiopeia~A radio flux
density relative to Cygnus~A at  151.5~MHz
${(S_{\rm CasA}/S_{\rm CygA})}_{151.5\,{\rm MHz}} \equiv r_{151.5}(t)$.
The straight line of Fig.\,2
$$
r_{151.5}(t) = m_{151.5}(t-\langle t\rangle_{151.5}) + c_{151.5},$$
where

$\langle t\rangle_{151.5} =1991.8$ is the mean epoch of
measurements at 151.5~MHz,

$m_{151.5} = dr_{151.5}(t) /dt =
-(8.779 \pm 0.851)\cdot10^{-3}\,\hbox{year}^{-1}$,

$c_{151.5} = r_{151.5}(\langle t\rangle_{151.5}) =
0.996 \pm 0.008$,\\
shows a weighted least-squares fit. The average
value of the secular decrease rate of the Cassiopeia~A radio emission
over the time interval 1980--2002 is equal to
\begin{equation}\label{f2}
d_{151.5\,{\rm MHz}}(1980\!-\!2002) =
100\cdot m_{151.5}/ c_{151.5} =- (0.88 \pm 0.09)\%\,\hbox{year}^{-1}.
\end{equation}

\begin{figure}[h]

\begin{minipage}[t]{50mm}
\includegraphics[width=5.5cm]{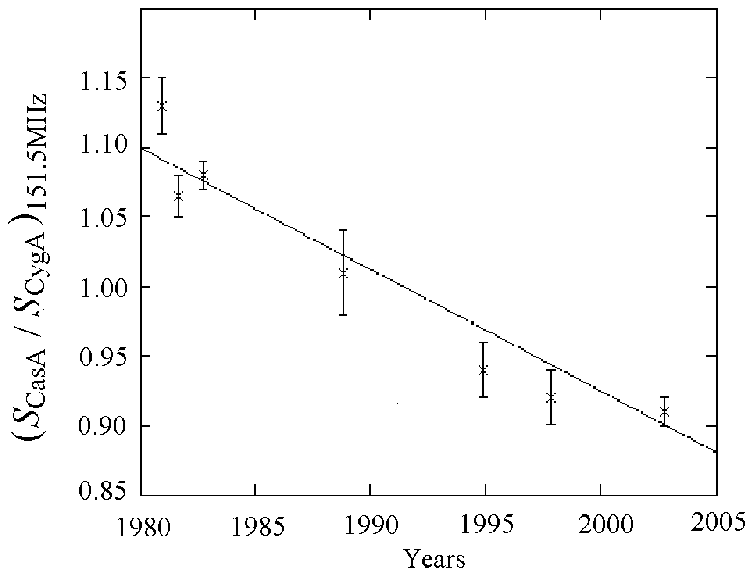}
\caption[]{Flux density of the Cassiopeia~A radio emission
relative to that of Cygnus~A at 151.5~MHz according to the
measurements at ``Staraya Pustyn$'$'' versus time. }
\end{minipage}
\hspace{10mm}
\begin{minipage}[t]{50mm}

\includegraphics[width=5.5cm]{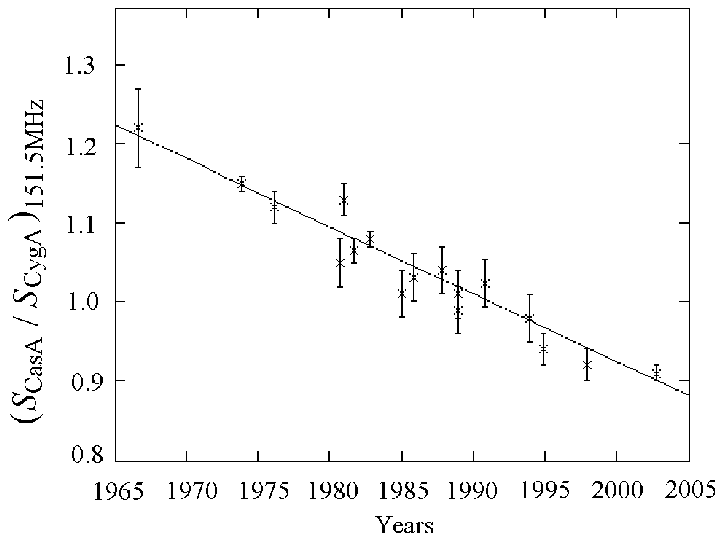}

\caption[]{Flux density of the Cassiopeia~A radio emission
relative to that of Cygnus~A at 151.5~MHz according to the
measurements at ``Staraya Pustyn$'$'' and the data of~[4--6]
versus time.}
\end{minipage}

\label{F2-3}
\end{figure}

The measurement results of $(S_{\rm CasA}/S_{\rm CygA})$
of other authors at 151 and 152~MHz are available in the
literature for the interval 1966--1993~[4--6]. Fig.\,3
shows the results of all known measurements at $\approx151.5$~MHz
(17 epochs altogether) including ``Staraya Pustyn$'$" data
together with a straight line of a weighted least-squares fit.
The average value of the secular decrease rate of the
Cassiopeia~A radio emission over the time interval  1966--2002
is equal to
\begin{equation}\label{f3}
d_{151.5\,{\rm MHz}}(1966\!-\!2002)=-(0.81 \pm 0.04)\%\,
\hbox{year}^{-1},
\end{equation}
that coincides within the limits of errors with the value~(2)
obtained at the observatory ``Staraya Pustyn$'$" over the interval
1980--2002.

\section{
Discussion
}
\label{sec6}

We can see from Fig.\,1 that the decline of Cassiopeia~A flux density with time
is not uniform. For example, in the beginning of the 1980's, the decrease
was more rapid than at the end of this decade. In addition, in 1979--1980
Cassiopeia~A flux density even increased.

It is interesting to compare the obtained values of the secular
decrease rate of Cassiopeia~A radio emission at 151.5~MHz~(2) and
927~MHz~(1) with the values of $d$ followed from the empirical
formula given in~[7]
\begin{equation}\label{f4}
d_\nu(\%\,\hbox{year}^{-1}) = -(0.97 \pm 0.04) +
(0.30 \pm 0.04) {\rm log}_{10}(\nu/1000\,{\rm MHz}).
\end{equation}

\begin{figure}[h]
\begin{minipage}{50mm}
\includegraphics[width=6cm]{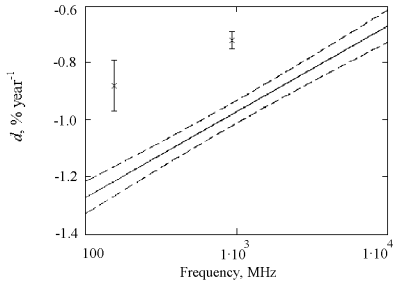}
\end{minipage}
\hspace{5mm}
\begin{minipage}{50mm}
\caption[]{
Comparison of the values of the Cassiopeia~A secular
decrease rate according to formula~(4) from~[7] (solid curve
gives the values of $d$ without account of errors, the dash
ones do that with the account of errors) and the values~(1)
and~(2) at 927 and 151.5~MHz, respectively.
}
\end{minipage}

\label{F4}
\end{figure}

Figure~4 shows $d(\nu)$ according to formula~(4) (with an account
of the errors the values of $d$ according to formula~(4) from~[7]
lie between dash lines in Fig.\,4) and the values of $d$~(1)
and~(2) according to our measurements. As seen from Fig.\,4
the values of $d$ obtained mainly by the measurements
in the last quarter of the 20-th century are substantially less by the
absolute value than the values of $d$ obtained by the measurements in the
third quarter of the 20-th century. This testifies to the
slowing down with time the decrease of Cassiopeia~A radio emission
(see also~[8]). At the same time the values of the
secular decrease rate at 151.5~MHz~(2) and 927~MHz~(1) do not
contradict to the conclusion on the secular flattening of the Cassiopeia~A
radio spectrum made in~[7, 9, 10].

\section{
Conclusion
}
\label{sec7}

As a result of long-term (1977--2002) measurements of the radio flux
density of Cassiopeia~A relative to Cygnus~A at 927 and 151.5~MHz using a single
radio telescope (radio interferometer) at a given frequency, we have
found the following mean secular decrease rates of the radio flux of
Cassiopeia~A:
$$
d_{927\,{\rm MHz}}(1977\!-\!2002)=-(0.72\pm 0.03)\%\,\hbox{year}^{-1},$$
$$
d_{151.5\,{\rm MHz}}(1980\!-\!2002)=-(0.88 \pm 0.09)\%\,
\hbox{year}^{-1}.$$
Our values of $d$ obtained by the measurements during the last 25 years are
substantially less by the absolute value than the values

$d(151.5\,{\rm MHz}) =- (1.22\pm 0.05)\%\,{\rm year}^{-1}$\\
and

$d(927\,{\rm MHz})=- (0.98\pm 0.04)\%\,\hbox{year}^{-1}$\\
which follow from the formula~(4)~[7]. This indicates to the slowdown of
Cassiopeia~A radio emission secular decrease.

In addition to this large scale time variation of Cassiopeia~A flux density
the observations have also shown a small scale (a few years) time
variations over the smooth secular decrease.

\section*{
Acknowledgment
}
\label{sec8}
This work has been supported by the International Science and
Technology Center under the ISTC project  No.\,729.


\begin{thebibliography}{9}
\bibitem{1}
                 Vinyajkin~E.\,N.,         Razin~V.\,A.:
Australian Journal of Physics  \textbf{32}, 93 (1979)

\bibitem{2}
                Vinyajkin~E.\,N.:
Astronomical and Astrophysical Transactions \textbf{11}, 325 (1996)

\bibitem{3}
                Vinyajkin~E.\,N.:
Astrophys.\ and Space Sci. \textbf{252}, 249 (1997)

\bibitem{4}
        Parker E.\,A.:
Mon.\ Not.\ R.\ Astr.\ Soc.  \textbf{138}, 407 (1968)

\bibitem{5}
        Read P.\,L.:
Mon.\ Not.\ R.\ Astr.\ Soc. \textbf{178}, 259 (1977)

\bibitem{6}
        Agafonov M.\,I.:
Astron.\ Astrophys. \textbf{306}, 578  (1996)

\bibitem{7}
        Baars J.\,W.\,M., Genzel~R., Pauliny-Toth~I.\,I.\,K.,
Witzel~A.:
Astron.\ Astrophys. \textbf{61}, 99 (1977)

\bibitem{8}
        Reichart D.\,E., Stephens A.\,W.:
Astrophys.\ J. \textbf{537}, 904  (2000)

\bibitem{9}
        Dent W.\,A., Aller H.\,D., Olsen E.\,T.:
Astrophys.\ J. \textbf{188}, L11 (1974)

\bibitem{10}
        Vinyajkin E.\,N., Razin V.\,A., Khrulev V.\,V.:
Soviet Astronomy Letters \textbf{6}, 324 (1980)
\end{thebibliography}
\end{document}